\documentclass[
aps,
prl,
 amsmath,amssymb,
preprint,
longbibliography,
superscriptaddress,
]{revtex4-1}

\usepackage{graphicx}
\usepackage{dcolumn}
\usepackage{bm}

\usepackage[T1]{fontenc} 	
\usepackage[utf8]{inputenc} 
\usepackage[dvipsnames]{xcolor}

\graphicspath{ {images/} }	
\usepackage{units}

\usepackage{fancyhdr}
\usepackage[acronym]{glossaries}
	\newacronym{sdof}{SDOF}{Single Degree of Freedom}
	\newacronym{mdof}{MDOF}{Multiple Degree of Freedom}
	\newacronym{dof}{DOF}{Degree of Freedom}
	\newacronym{dofs}{DOFs}{Degrees of Freedom}
	\newacronym{mbs}{MBS}{Multibody Simulation}
	\newacronym{fem}{FEM}{Finite Element Method}
	\newacronym{fea}{FEA}{Finite Element Analysis}
	\newacronym{fe}{FE}{Finite Element}
	\newacronym{ode}{ODE}{Ordinary Differential Equation}
	\newacronym{cae}{CAE}{Computer-aided Engineering}
	\newacronym{ema}{EMA}{Experimental Modal Analysis}
	\newacronym{oma}{OMA}{Operational Modal Analysis}
	\newacronym{soe}{SOE}{System of Equations}
	\newacronym{rom}{ROM}{Reduced Order Modeling}
	\newacronym{kp}{KP}{Keypoint/s}
	\newacronym{msd}{MSD}{Mode Shape Deviation}
	\newacronym{biw}{BIW}{Body in White}
	\newacronym{rbm}{RBM}{Rigid Body Mode/s}
	\newacronym{cad}{CAD}{Computer-aided Design}
	\newacronym{spl}{SPL}{Sound Pressure Level}
	\newacronym{gmm}{GMM}{Gaussian Mixture Model}
	\newacronym{dft}{DFT}{Discrete Fourier Transform}
	\newacronym{fft}{FFT}{Fast Fourier Transform}
	\newacronym{hvac}{HVAC}{Heating, Ventilation and Air Conditioning}
	\newacronym{nc}{NC}{Noise Criterion}
	\newacronym{ncb}{NCB}{Balanced Noise Criterion}
	\newacronym{rc}{RC}{Room Criterion}
	\newacronym{pdf}{PDF}{Probability Density Function}
	\newacronym{aic}{AIC}{Akaike Information Criterion}
	\newacronym{bic}{BIC}{Bayesian Information Criterion}
	\newacronym{min1s}{min1s}{Minimum \unit[1]{s}-Averaged Level}
	\newacronym{amm}{AMM}{Acoustic Metamaterials}
	\newacronym{bem}{BEM}{Boundary Element Method}
    \newacronym{wc}{WC}{Willis Coupling}
    \newacronym{ama}{AMA}{Acoustic Meta-Atom}
    \newacronym{hr}{HR}{Helmholtz Resonator}
    \newacronym{pla}{PLA}{Polylactic Acid}

\begin{document}


\title{
Acoustic meta-atom with maximum Willis coupling}

\author{Anton Melnikov}
\email{anton.melnikov@tum.de}
\affiliation{
Vibroacoustics of Vehicles and Machines\\
Technical University of Munich, Germany}
\affiliation{
	SBS Bühnentechnik GmbH\\
	Dresden, Germany}
\affiliation{
	Centre for Audio, Acoustics and Vibration\\
	University of Technology Sydney, Australia}
\affiliation{
	School of Engineering and Information Technology\\
	University of New South Wales, Canberra, Australia}

\author{Li Quan}
\affiliation{
Department of Electrical and Computer Engineering\\
The University of Texas at Austin, Austin, Texas 78712, USA}
 
\author{Sebastian Oberst}
\affiliation{
Centre for Audio, Acoustics and Vibration\\
University of Technology Sydney, Australia}

\author{Andrea Alù}
\affiliation{
Metamaterials and Plasmonics Research Laboratory\\
The University of Texas at Austin, USA}
\affiliation{Photonics Initiative, Advanced Science Research Center\\
City University of New York, USA}
 
\author{Steffen Marburg}
\affiliation{
Vibroacoustics of Vehicles and Machines\\
Technical University of Munich, Germany}

\author{David Powell}
\email{david.powell@adfa.edu.au}
\affiliation{
School of Engineering and Information Technology\\
University of New South Wales, Canberra, Australia}
\date{\today}
            
\begin{abstract}
Acoustic metamaterials are structures with exotic acoustic properties, having promising applications in acoustic beam steering, focusing, impedance matching, absorption and isolation. Recent work has shown that the efficiency of many acoustic metamaterials can be enhanced by controlling an additional parameter known as Willis coupling, which  is analogous to bianisotropy in electromagnetic metamaterials. The magnitude of Willis coupling in an acoustic meta-atom has been shown theoretically to have an upper limit, however the feasibility of reaching this limit has not been experimentally investigated. Here we introduce a meta-atom with Willis coupling which closely approaches this theoretical limit, that is much simpler and less prone to thermo-viscous losses than previously reported structures.
We perform two-dimensional experiments to measure the strong Willis coupling, supported by numerical calculations.
Our meta-atom geometry is readily modeled analytically, enabling the strength of Willis coupling and its peak frequency to be easily controlled.
Together with its ease of fabrication, this will facilitate the design of future high efficiency acoustic devices. 
\end{abstract}

\maketitle

Acoustic metamaterials \cite{cummer_controlling_2016,koo_acoustic_2016} have demonstrated unique material properties which do not exist naturally, such as negative bulk modulus \cite{fang_ultrasonic_2006} and negative dynamic mass density \cite{liu_locally_2000}.
These properties have enabled the development of acoustic superlenses \cite{ambati_surface_2007, zhang_focusing_2009, torrent_acoustic_2007}, barriers \cite{elford_matryoshka_2011}, cloaking devices \cite{zhang_broadband_2011} and the enhancement of non-linear effects \cite{quan_quasi-phase-matched_2012}. Metamaterials are typically arrays of sub-wavelength structures, known as meta-atoms, with geometry engineered to control their dynamic mass and stiffness. It has recently been shown that more efficient metamaterial designs can be created by incorporating an additional degree of freedom, represented by the Willis coupling parameter.

Willis coupling is a term in the acoustic and elastic constitutive relations that couples potential and kinetic energy \cite{willis_variational_1981,willis_nonlocal_1985,willis_effective_2011}, analogous to the bianisotropy parameter in electromagnetism \cite{cheng_covariant_1968,kong_theorems_1972}. The Willis coupling and bianisotropy parameters can only be non-zero in structures which lack mirror symmetry about one of their major axes, and their inclusion into the constitutive relations has been shown to resolve violations of causality and passivity in metamaterial homogenization \cite{muhlestein_reciprocity_2016,alu_restoring_2011}. Recent work has demonstrated that the incorporation of Willis coupling or bianisotropy into metamaterial structures of sub-wavelength thickness, known as metasurfaces, can improve their efficiency when refracting at large angles \cite{li_systematic_2018, quan_maximum_2018, radi_metagratings:_2017}. While bianisotropy has been demonstrated and engineered in a wide range of electromagnetic meta-atom designs \cite{AsadchyBianisotropicmetasurfacesphysics2018a}, approaches for controlling the degree of Willis coupling in acoustic meta-atoms are not well-established.

Recently, a bound on the maximum value of the Willis coupling parameter was derived, based on the conservation of energy \cite{quan_maximum_2018}. It was shown how meta-atoms can be designed to reach this theoretical bound, using space-coiling structures with long meander-line channels \cite{LiAcousticfocusingcoiling2012, cheng_ultra-sparse_2015, lu_realization_2017, quan_maximum_2018}.
While such structures are advantageous for achieving resonance in a sub-wavelength volume, they are difficult to manufacture reproducibly, typically requiring additive manufacturing techniques.
Moreover, their channel widths are often comparable to the viscous and thermal boundary layer thickness, and their channel lengths are of the order of the wavelength, leading to high thermo-viscous losses, and a significant reduction in scattering efficiency \cite{attenborough_acoustical_1983, stinson_propagation_1991, jordaan_measuring_2018}. It was shown numerically in Ref.~\cite{quan_maximum_2018} that the thermo-viscous losses in space-coiling meta-atoms can reduce their Willis coupling magnitude to be significantly lower than the theoretical bound.

Experimental evidence of Willis coupling has been reported in both one-dimensional  \cite{koo_acoustic_2016,muhlestein_experimental_2017}, and two-dimensional metamaterial structures \cite{li_systematic_2018}. However, the theoretical limit on the strength of Willis coupling has not been tested experimentally, and it remains unknown how closely this limit may be approached in practice. To resolve this question we propose an acoustic meta-atom, which is designed to realize maximum Willis coupling and to minimize thermo-viscous losses. Experimental results obtained from a fabricated sample are compared to numerical calculations, showing good agreement of the resonant frequency and lineshape, with a reduction in magnitude that we attribute primarily to the compliance of the 3D printed material. The simplicity of the structure enables us to present an analytical model for its polarizability, showing how the Willis coupling can be tailored to have any value between zero and the theoretical bound.

\section{Results}

Acoustic wave interaction with a meta-atom is conveniently described by its polarizability tensor. Due to the sub-wavelength size of meta-atoms, their scattering is dominated by monopole and dipole moments, and their polarizability can be written as
\begin{equation}\label{eq:MDD}
\begin{bmatrix}
M \\ \mathbf{D}
\end{bmatrix}
= \bm{\alpha}
\begin{bmatrix}
\breve{p}^{\textrm{inc}} \\ \breve{\mathbf{v}}^{\textrm{inc}}
\end{bmatrix}
=
\begin{bmatrix}
\alpha^{pp}			& \bm{\alpha}^{pv}	\\
\bm{\alpha}^{vp}	&\bm{\alpha}^{vv}	\\
\end{bmatrix}
\begin{bmatrix}
\breve{p}^{\textrm{inc}} \\ \breve{\mathbf{v}}^{\textrm{inc}}
\end{bmatrix},
\end{equation}
where $M$ is the scalar monopole moment, $\mathbf{D}$ is the vector dipole moment, $\bm{\alpha}$ is the polarizability tensor, $\breve{p}^{\textrm{inc}}$ and $\breve{\mathbf{v}}^{\textrm{inc}}$ are the incident pressure and the velocity at the center of the meta-atom \cite{quan_maximum_2018}.
The off-diagonal terms $\bm{\alpha}^{pv}$ and $\bm{\alpha}^{vp}$ represent the Willis coupling between the dipolar and monopolar responses. 

\subsection{Meta-atom design}

We show how the reported structures exhibiting Willis coupling \cite{koo_acoustic_2016,muhlestein_experimental_2017,quan_maximum_2018} can be replaced by a simpler, and more reproducible structure, by avoiding thin channels and large areas of fluid-structure interfaces. The structure can be tuned to achieve any value of Willis coupling up to the theoretical bounds and can be readily analyzed using a closed form analytical solution which assists in understanding the physical mechanisms behind its operation.

\begin{figure}[hbtp]
	\includegraphics[width=.5\linewidth]{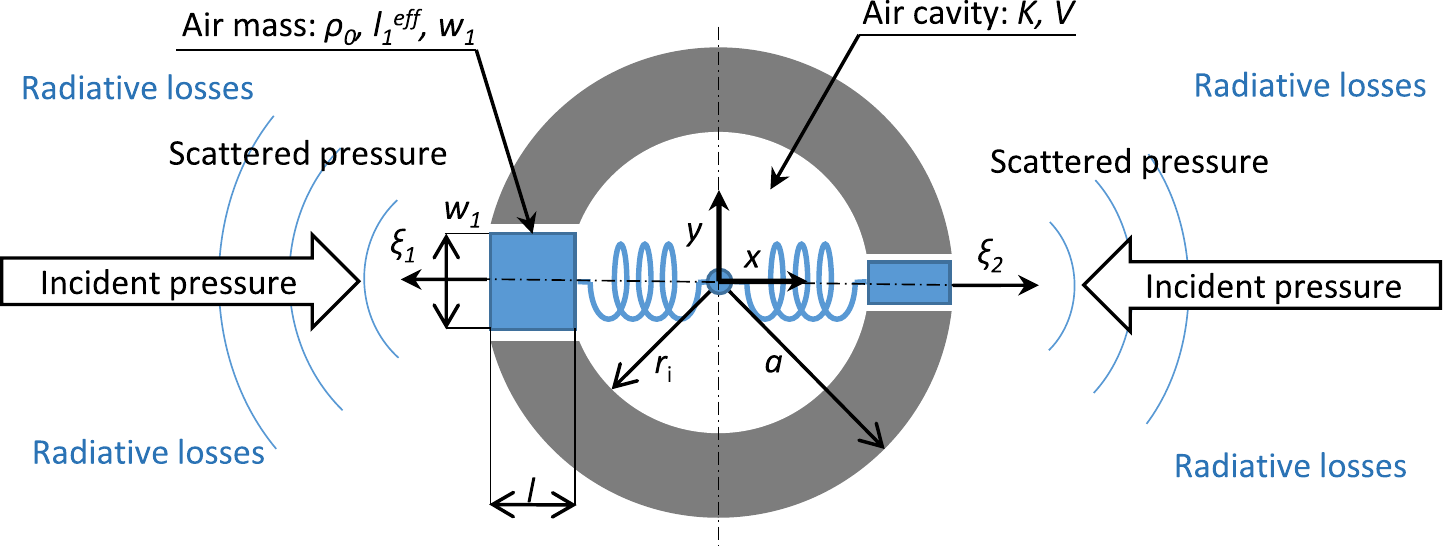}
	\caption{
		Geometrical shape and dimensions of meta-atom, where $a$ is the cylinder radius and $r_i$ is the cavity radius.
		The neck widths $w_n$ are in general different for each aperture, the neck length $l$ is common to all apertures and the cavity volume $V = \pi r_i^2$.
		\label{fig:shape}}
\end{figure}

Achieving a strong acoustic polarizability in a small volume requires a resonant structure.
Inspired by resonant sonic crystals \cite{elford_matryoshka_2011} and Helmholtz resonators with multiple apertures \cite{dosch_radiative_2016,crow_experimental_2015, kinsler_fundamentals_2000}, our novel meta-atom design is presented in Fig.~\ref{fig:shape}.
This meta-atom exhibits Willis coupling due to the asymmetrical neck openings.
The air within each  neck is treated as an incompressible mass, while the air in the cavity acts as a spring, together creating an oscillating system excited by an incident acoustic wave.
Since the oscillation occurs in the fluid domain only, this avoids wave coupling through fluid-structure interfaces.
Since the structure avoids long and thin channels, thermo-viscous losses are expected to be greatly reduced.

Peak Willis coupling is expected to occur close to the Helmholtz resonator's eigenfrequency, which is dependent on the air mass moving within the apertures and the inner cavity volume.
In two dimensions (2D), the moving mass is determined by the aperture cross-sections $A_n = w_n$ and the neck length $l$, which is equal for all apertures due to the inner and outer cylindrical boundaries being concentric.
The 2D-volume $V = \pi r_\textrm{i}^2$ is determined by the inner radius $r_\textrm{i} = a - l$, where $a$ is the outer radius.
Following this, the eigenfrequency for $N$ apertures can be approximated as \cite{supplementary}
\begin{equation}\label{eq:omega_0}
\omega_0 = \frac{c}{r_\mathrm{i}} \sqrt{\frac{\sum_{n=1}^N w_n}{\pi l}}.
\end{equation}
Generally an arbitrary number of apertures may be included. However for controlling Willis coupling along one axis,  two oppositely arranged apertures are sufficient, as illustrated in Fig.~\ref{fig:shape}. The apertures can be used to adjust the resonant frequency and the level of Willis coupling.
The maximum Willis coupling magnitude is exhibited when the structure shown in Fig.~\ref{fig:shape} becomes maximally asymmetric, consistent with the single aperture meta-atoms shown in Ref.~\cite{quan_maximum_2018}.

\subsection{Experimental verification}

\begin{figure*}[hbtp]
	\begin{tabular}{lll}
    	\includegraphics[width=.212\textwidth]{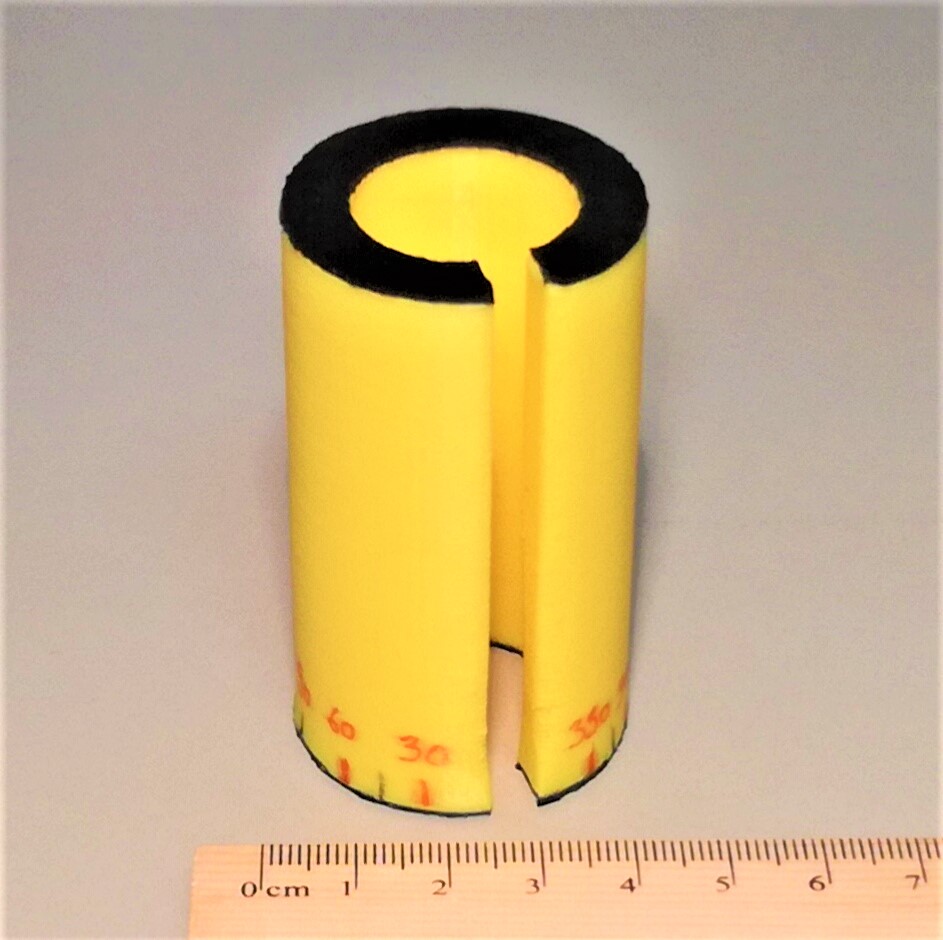} & 
    	\includegraphics[width=.385\textwidth]{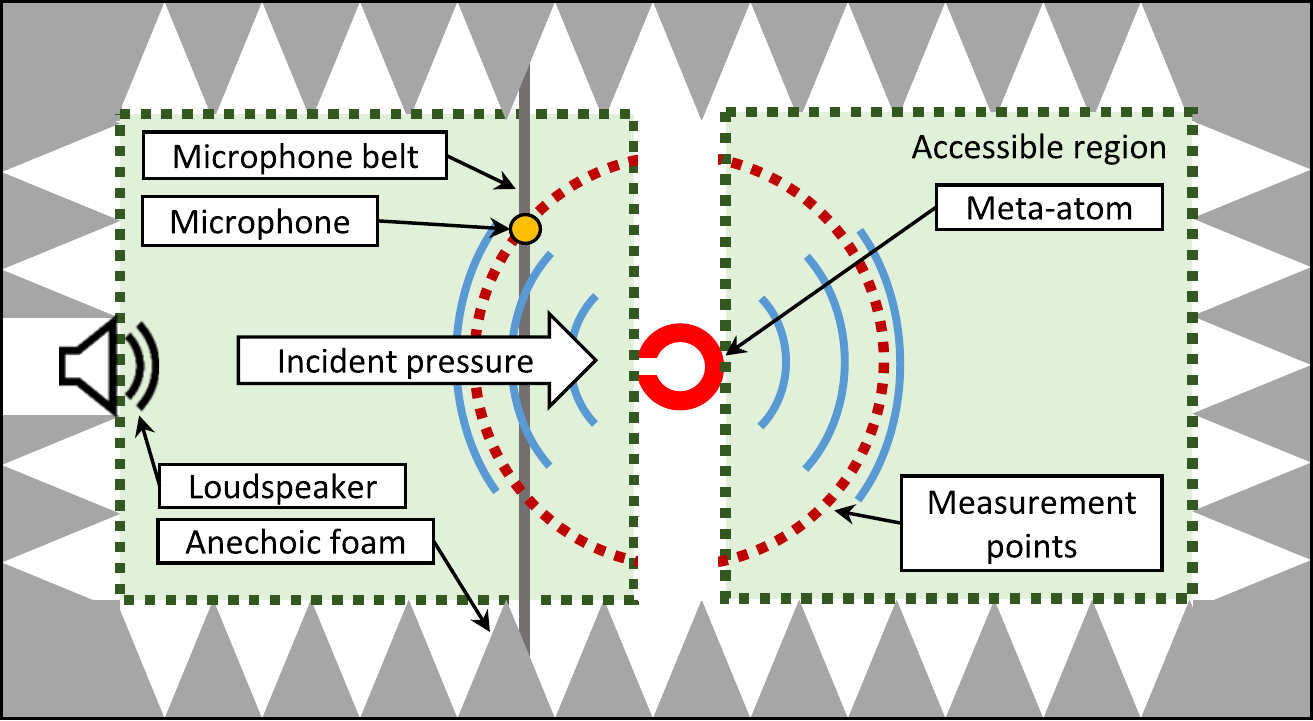} &
    	\includegraphics[width=.35\textwidth]{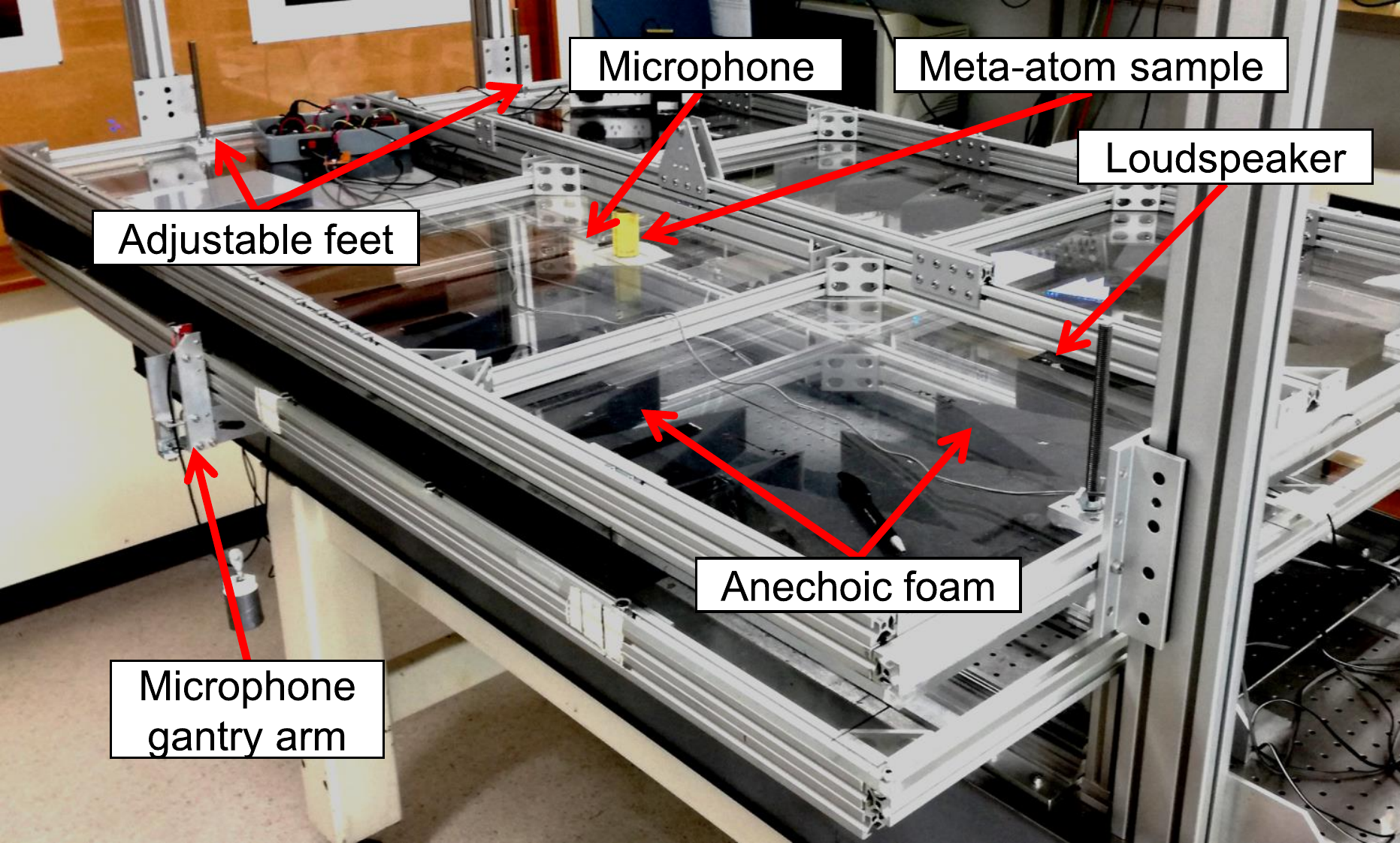}
    	\\[-2.5em]
        (a) & (b) & (c) \\[.7em]
    \end{tabular}
    \includegraphics[width=1\textwidth]{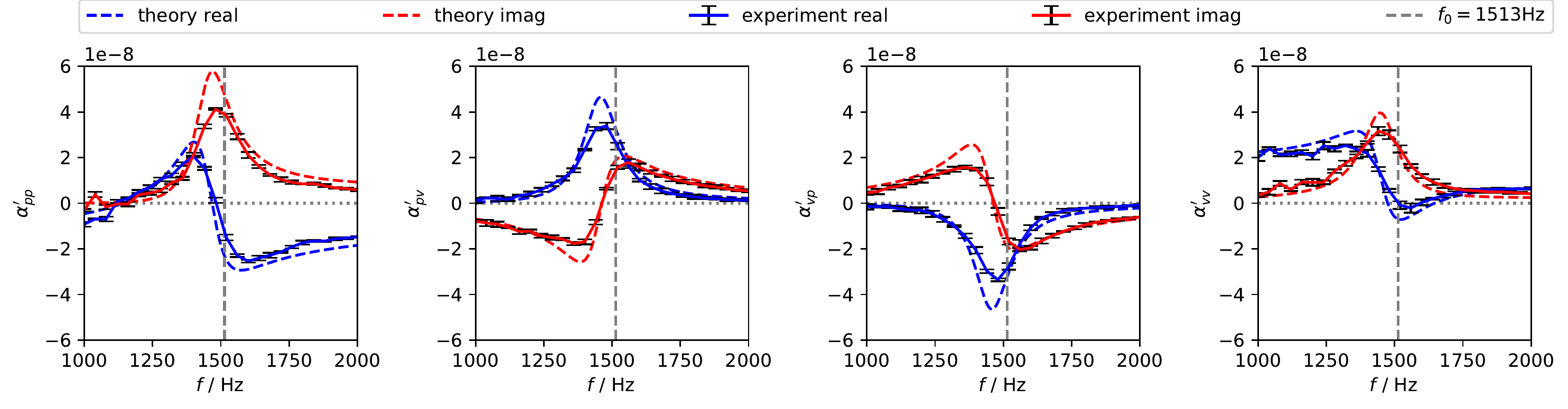}
    \\[-4em]
    \begin{tabular}{llll}
        \hspace{1cm}(d)\hspace{2.5cm} &
        \hspace{1cm}(e)\hspace{2.5cm} &
        \hspace{1cm}(f)\hspace{2.5cm} &
        \hspace{1cm}(g)\hspace{2.5cm} \\[3em]
    \end{tabular}
    \caption{
		(a) 3D-printed sample: single aperture meta-atom with $a = \unit[19]{mm}$, $r_\mathrm{i} = \unit[12]{mm}$, $w = \unit[5.5]{mm}$ and $h = \unit[66]{mm}$.
        (b, c): Schematic and photograph of experimental set-up.
        (d-g): Theoretically predicted and experimentally determined components of the normalized polarizability tensor. The error bars show the standard deviation resulting from the least squares fit over six incident angles.
		\label{fig:experiment_full}}
\end{figure*}

To experimentally demonstrate  Willis coupling in the presented meta-atom, its polarizability is determined in a 2D experiment.
A single aperture meta-atom providing maximum asymmetry is manufactured and investigated (see Fig.~\ref{fig:experiment_full}(a)). The incident and scattered pressure fields are measured in a 2D parallel-plate waveguide \cite{jordaan_measuring_2018} (see Fig.~\ref{fig:experiment_full}(b)) with subsequent extraction of the polarizability tensor as detailed in the Supplementary Material \cite{supplementary}.

\begin{figure}[hbtp]
	\includegraphics[width=.4\linewidth]{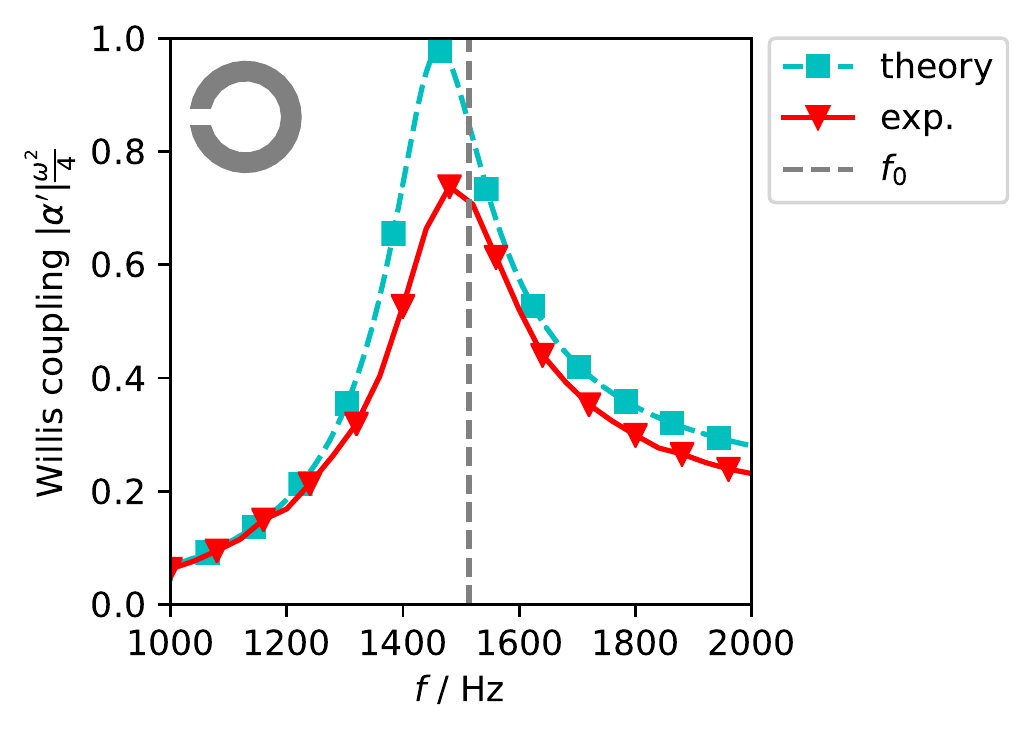}
	\caption{
		Experimentally determined and theoretically predicted Willis coupling magnitude for a single aperture meta-atom, normalized to the theoretical bound $4 \omega^{-2}$.
		\label{fig:max_willis}}
\end{figure}

The polarizability tensor defined in Eq.~\eqref{eq:MDD} has elements with different units and values differing by many orders of magnitude. Therefore it is convenient to introduce the  normalized polarizability tensor $\bm{\alpha}^{\prime}$, the elements of which are shown for our meta-atom structure in Fig.~\ref{fig:experiment_full}(d-g).
The normalized values are defined as
$\alpha^{\prime}_{pp} = -2\alpha^{pp}$,
$\alpha^{\prime}_{pv} = \frac{-\sqrt{2}}{\rho c}\alpha^{pv}$,
$\alpha^{\prime}_{vp} = ik\sqrt{2}\alpha^{vp}$, and
$\alpha^{\prime}_{vv} = \frac{ik}{\rho c}\alpha^{vv}$ \cite{quan_maximum_2018}.
As required by reciprocity\cite{quan_maximum_2018}, the tensor satisfies $\bm{\alpha}^{\prime} = \bm{\alpha}^{\prime T-}$, since the off-diagonal terms are equal to each other with a sign reversal (see Fig.~\ref{fig:experiment_full}(e) and \ref{fig:experiment_full}(f)).
Furthermore, all polarizability components show strong peaks close to the eigenfrequency predicted by Eq.~\eqref{eq:omega_0}.
To investigate the magnitude of the meta-atom's Willis coupling, Fig.~\ref{fig:max_willis} shows the average of $\left|\alpha^{\prime}_{pv}\right|$ and $\left|\alpha^{\prime}_{vp}\right|$ determined experimentally (red solid line) and analytically (cyan dashed line).
Here, the peak Willis coupling at \unit[1480]{Hz} achieves \unit[74]{\%} of the theoretical bound $4/\omega^{2}$.

\subsection{Polarizability theory}

To show how the Willis coupling and other polarizability components can be tailored by adjusting the meta-atom geometry, we develop a polarizability theory. The predictions of this theory are shown in Figs.~\ref{fig:experiment_full} and \ref{fig:max_willis}. It is based on multiple aperture Helmholtz resonator dynamics for the outward directed particle displacement $\xi_n$ within each aperture
\begin{equation}\label{eq:hwr}
i\omega^3 c^{\mathrm{rad}}_n \frac{\rho_0A_n}{c}\xi_{n} + \omega^2 \rho_0 l_n^{\mathrm{eff}} \xi_n - \frac{K}{V}\sum_mA_m\xi_m = p^{\mathrm{ext}}_{n}.
\end{equation}
Here $p^{\mathrm{ext}}_{n}$ is the external pressure at aperture $n$, $\rho_0$ is the medium density, $c$ is the speed of sound, $A_n = w_n$ is the cross-section of the aperture, $K = c^2 \rho_0$ is the bulk modulus, $V$ is the inner cavity volume, $l_n^{\mathrm{eff}} = l + c_n^{\mathrm{eff}}$ is the effective neck length (including a radiative correction) and $c^{\mathrm{rad}}_n$ is the radiative loss coefficient (see Fig.~\ref{fig:shape}). The summation over $m$ accounts for coupling between apertures via compression of the center cavity. Considering all apertures results in a matrix equation $\mathbf{K}_{\textrm{eq}} \bm{\xi} = \mathbf{p}^{\textrm{ext}}$, where $\mathbf{K}_{\textrm{eq}}$ is the dynamic stiffness matrix given by Eq.~(S38) of the Supplementary Material. This matrix allows the displacement to be solved for an arbitrary incident pressure field and hence, the contribution of the oscillating air masses (see Fig.~\ref{fig:shape}) to the scattering of the meta-atom. See the Supplementary Material \cite{supplementary} for full details of the derivation.

We consider now the shape illustrated in Fig.~\ref{fig:shape}.
This restricts the Willis coupling to a single axis and simplifies the problem to a $2 \times 2$ matrix.
In this case the resonator polarizability tensor $\bm{\alpha}^{\mathrm{res}}$ is obtained as
\begin{equation}\label{eq:ares2}
\bm{\alpha}^{\mathrm{res}} = 
\rho_0
\begin{bmatrix}
A_1 & A_2 \\
x_1 A_1 & x_2 A_2
\end{bmatrix}
\mathbf{K}_{\mathrm{eq}}^{-1}
\begin{bmatrix}
1 & -i a \rho_0 \omega \\
1 & i a \rho_0 \omega
\end{bmatrix},
\end{equation}
The ratio between aperture widths $\frac{w_1}{w_2} = \frac{A_1}{A_2}$ determines the strength of Willis coupling and can be used to optimize the structure.
Expanding Eq.~\eqref{eq:ares2} for a single aperture meta-atom with $w_2=0$ leads to the following expression for the polarizability
\begin{equation}\label{eq:ares1}
\bm{\alpha}^{\mathrm{res}} = \frac{\rho_0
	\begin{bmatrix}
	A_1		& -i a \rho_0 \omega A_1\\
	x_1 A_1	& -i a \rho_0 \omega x_1 A_1\\
	\end{bmatrix}}{i\omega^3 c^{\mathrm{rad}} \frac{\rho_0 A_1}{\pi c} + \omega^2 \rho_0 l_1^{\mathrm{eff}} - \frac{K}{V} A_1}.
\end{equation}
The dynamic stiffness matrix becomes a scalar equation with the remaining projection matrix being singular. The singularity arises because both the monopole and dipole moments are determined from the air movement within a single aperture, but it presents no computational difficulties, since the inverse of the polarizability tensor is not required.

As the Helmholtz resonator is embedded within a cylinder (see Fig.~\ref{fig:shape}), there is an additional influence on the meta-atom polarizability due to the background scattering from the cylinder. Considering only the dipole and monopole terms, the polarizability tensor of a cylinder of radius $a$ is \cite{supplementary}
\begin{equation}\label{eq:acylxy}
\bm{\alpha}^{\textrm{cyl}} =
\begin{bmatrix}
\frac{4}{i k^2 c^2} \frac{J_1(ka)}{ H_1^{(1)}(ka)} & 0\\
0 & \frac{8 \rho_0}{k^3 c} \frac{J_1^{\prime}(ka)}{ H_1^{(1)\prime}(ka)}\\
\end{bmatrix},
\end{equation}
where $J_n$ is a Bessel function,  $H_n^{(1)}$ is a Hankel function of the first kind and $k=\omega/c$ is the wavenumber. As expected for a symmetrical geometry, the off-diagonal terms corresponding to Willis coupling are zero. Since the resonator and cylinder are superimposed, they influence the effective incident fields of each other through an additional scattered term:
\begin{equation}\label{eq:pinc}
\begin{aligned}
\breve{p}^{\mathrm{inc}}_{\mathrm{cyl}} = \breve{p}^{\mathrm{inc}} + \breve{p}^{\mathrm{scat}}_{\mathrm{res}}\\
\breve{p}^{\mathrm{inc}}_{\mathrm{res}} = \breve{p}^{\mathrm{inc}} + \breve{p}^{\mathrm{scat}}_{\mathrm{cyl}}.
\end{aligned}
\end{equation}
This results in a coupled formulation for the monopole and dipole moments of the cylinder and the Helmholtz resonator as
\begin{equation}\label{eq:coupling}
\begin{bmatrix}
M^{\mathrm{cyl}} \\ D^{\mathrm{cyl}} \\ M^{\mathrm{res}} \\ D^{\mathrm{res}} \\
\end{bmatrix}
=
\begin{bmatrix}
\mathbf{I}						& -\bm{\alpha}^{\mathrm{cyl}}\mathbf{E}\\
-\bm{\alpha}^{\mathrm{res}}\mathbf{E}	& \mathbf{I}\\
\end{bmatrix}^{-1}
\begin{bmatrix}
\bm{\alpha}^{\mathrm{cyl}}\mathbf{u}^{\mathrm{inc}}\\
\bm{\alpha}^{\mathrm{res}}\mathbf{u}^{\mathrm{inc}}
\end{bmatrix}.
\end{equation}
where
$\mathbf{u}^{\mathrm{inc}} = \begin{bmatrix} \breve{p}^{\mathrm{inc}} & \breve{v}^{\mathrm{inc}}_{x} \end{bmatrix}^T$
is the incident field vector and
$
\mathbf{E} = \textrm{diag} \left( -\frac{ik^2c^2}{4}H_0^{(1)}(ka), -\frac{k^2 c}{4 a \rho_0} H _1^{(1)}(ka) \right)
$
represents the acoustic propagation from the cylinder to each of the apertures.
Finally, adding the monopole and dipole moments from Eq.~\eqref{eq:coupling} results in
\begin{equation}\label{eq:atot}
\begin{bmatrix}
M^{\mathrm{tot}} \\ D^{\mathrm{tot}}
\end{bmatrix}
=
\begin{bmatrix}
M^{\mathrm{cyl}} + M^{\mathrm{res}} \\ D^{\mathrm{cyl}} + D^{\mathrm{res}}
\end{bmatrix}
=
\bm{\alpha}^{\mathrm{tot}} \mathbf{u}^{\mathrm{inc}},
\end{equation}
which gives the total polarizability tensor $\bm{\alpha}^{\mathrm{tot}}$.
The components of $\bm{\alpha}^{\mathrm{tot}}$ for a single aperture meta-atom with $a = \unit[19]{mm}$, $r_\textrm{i} = \unit[12]{mm}$, $w = \unit[5.5]{mm}$, $c = \unit[343]{m/s}$, and $\rho_0 = \unit[1.2]{kg/m^3}$ are shown together with the experimental results in Fig.~\ref{fig:experiment_full}.

\section{Discussion}

\begin{figure}[hbtp]
	\includegraphics[width=.45\linewidth]{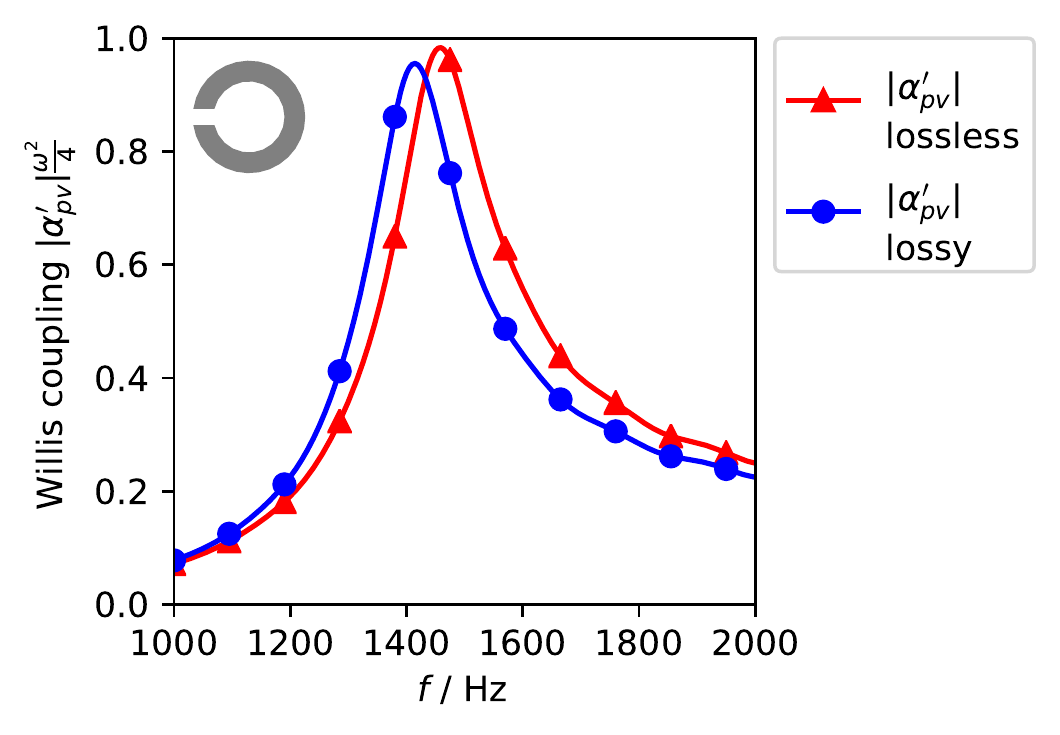}
	\caption{
		Numerical comparison of Willis coupling showing the influence of thermo-viscous losses in air, which cause a reduction in magnitude of only \unit[2.8]{\%}.
		\label{fig:lossy}}
\end{figure}

Figure \ref{fig:experiment_full} shows the measurement of strong Willis coupling in the proposed structure. It can be seen that the line-shape and resonant frequency are well described by our theoretical model. To illustrate how closely our meta-atom approaches the theoretical bound, the magnitude of the Willis coupling is plotted in Fig.~\ref{fig:max_willis}.
The difference between the experimental and theoretical values can be attributed to the thermo-viscous losses of air and the non-zero compliance of the meta-atom, which are not treated in the presented theory.
To investigate which of these mechanisms dominates the losses within the meta-atom, we first consider the thermo-viscous losses. These are modelled using the \gls{fem}, with the resulting Willis coupling shown in Fig.~\ref{fig:lossy}.
The results show that thermo-viscous losses lead to a reduction of only \unit[2.8]{\%} in the magnitude of Willis coupling. We note that this reduction is much less than that previously reported for space-coiling meta-atoms with thin channels (see Supplementary Material of Ref.~\cite{quan_maximum_2018}).

\begin{figure}[hbtp]
	\includegraphics[width=.45\linewidth]{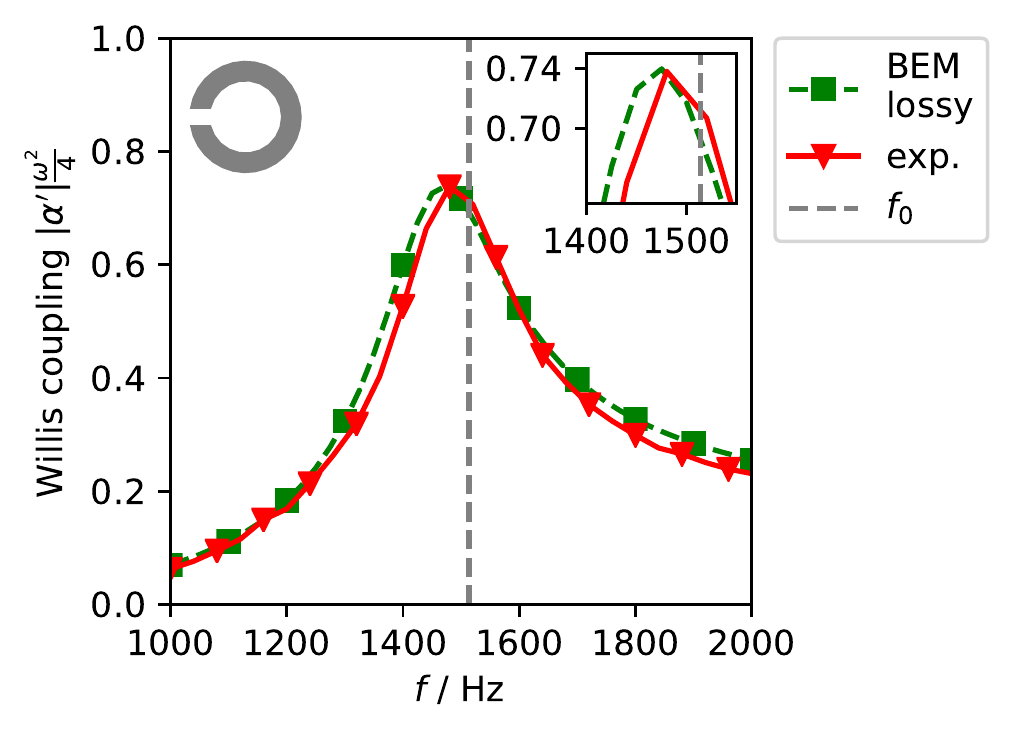}
	\caption{
		Willis coupling calculated numerically when incorporating a lossy surface admittance of $0.0057\left( \rho c \right)^{-1}$, reducing the peak value to match the experimental results.
		\label{fig:impedance}}
\end{figure}

Since viscous losses have a relatively minor influence on the magnitude of the Willis coupling, the reduction observed in the experiments can be attributed to the \gls{pla} from which the meta-atom is fabricated not having a perfectly hard boundary.
The \gls{pla} can be modelled as viscoelastic material, which can be represented within the \gls{bem} model through an equivalent surface admittance. The imaginary part of this admittance is expected to have negligible influence on the results.
Therefore, we use a purely real admittance of $0.0057\left( \rho c \right)^{-1}$, equivalent to an absorption of \unit[2.25]{\%}, consistent with experimentally measured values \cite{Chen_Study_on_Sound_2010}. As shown in Fig.~\ref{fig:impedance}, incorporating this surface admittance into the \gls{bem} model decreases the peak Willis coupling to approximately \unit[74]{\%} of the theoretical maximum, matching well with our experimental observations.

\begin{figure}[hbtp]
	\includegraphics[width=.45\linewidth]{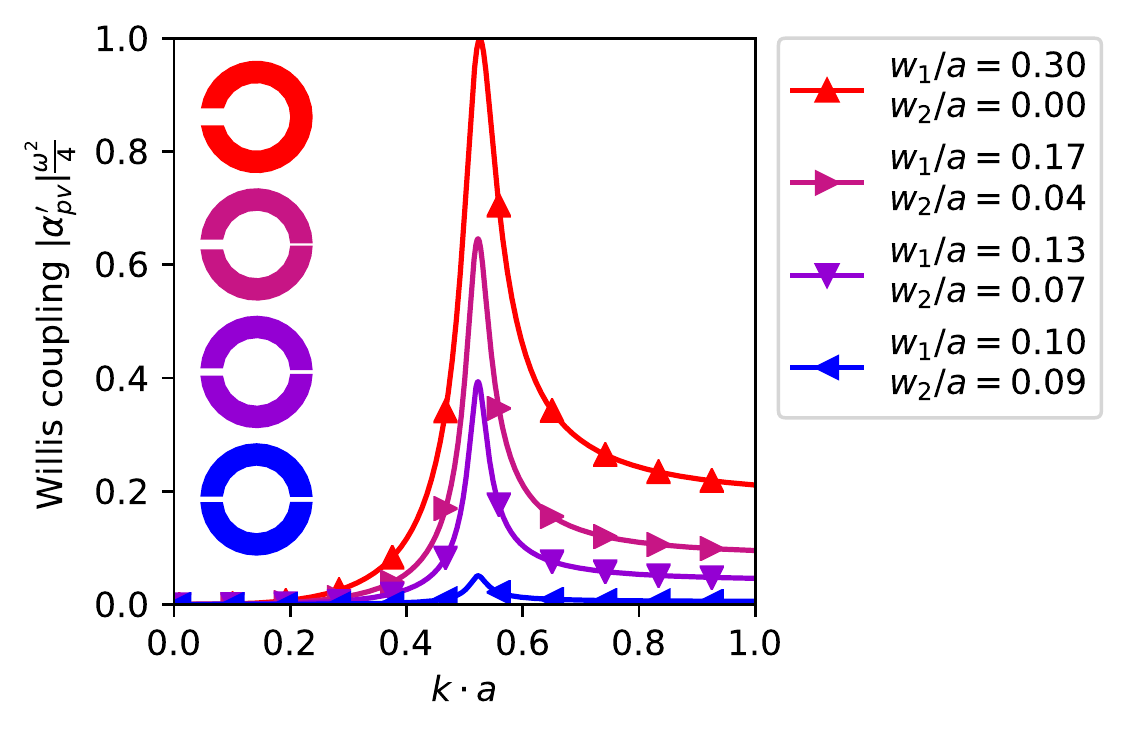}
	\caption{
		Willis coupling of four different meta-atom geometries, where $w_2$ is varied to control Willis coupling and $w_1$ is tuned to keep the peak frequency fixed.
		\label{fig:wc_tailoring_opt}}
\end{figure}

In addition to the experimentally demonstrated maximum Willis coupling, the structure presented in Fig.~\ref{fig:shape} can be tailored to have Willis coupling from zero up to the theoretical bound.
To demonstrate this property, normalized Willis coupling for four different parameter sets is shown in Fig.~\ref{fig:wc_tailoring_opt}. The red meta-atom on top illustrates the single aperture configuration similar to the experimentally investigated structure from  Fig.~\ref{fig:experiment_full}(a). Once a second aperture with neck width $w_2$ is introduced, Willis coupling is significantly reduced (Fig.~\ref{fig:wc_tailoring_opt} magenta and violet lines). This would result in a frequency shift, as expected from Eq.~\eqref{eq:omega_0}. To avoid this, $w_1$ is tuned to match the peak frequency of the single aperture case. When $w_2$ is further increased and the shape starts to converge to the symmetrical case (Fig.~\ref{fig:wc_tailoring_opt} blue, $\frac{w_1}{w_2} \approx 1.1$), the Willis coupling becomes very weak and disappears completely when $\frac{w_1}{w_2} = 1$. This behavior is of practical importance, since it allows tailoring of the Willis coupling to any desired values. A full parametric analysis of the influence of $w_1$ and $w_2$ on the resonant frequency and peak Willis coupling is presented in the Supplementary Material Fig.~S6 \cite{supplementary}.

\section{Conclusion}

We introduced and experimentally validated a novel meta-atom exhibiting strong Willis coupling.
The experiment revealed a Willis coupling magnitude reaching \unit[74]{\%} of the theoretical bound.
It was demonstrated numerically that this difference is dominated by losses within the meta-atom material (\gls{pla}). In this structure thermo-viscous losses in air are quite small, whereas they are much stronger in previously reported space-coiling meta-atoms.
Additionally, the simple shape of our structure facilitates manufacturing and enables accurate analytical modeling.
Combining the models of a Helmholtz resonator and a cylindrical scatterer, a theory was developed and shown to agree well with numerical simulations.
Since our structure enables Willis coupling to be tailored, this theory can be used to engineer Willis coupling for specific applications.



\section{Methods}

\subsection{Extraction of Polarizability Tensor}
The polarizability relationship given in Eq. \eqref{eq:MDD} is used to illustrate Willis coupling, where it appears as the off-diagonal terms $\bm{\alpha}^{pv}$ and $\bm{\alpha}^{vp}$. 
An extraction method is necessary to obtain the polarizability of a scatterer from an experiment or numerical model.
For simplicity, only the 2D case is considered.
We build on the method for extracting polarizability from highly symmetric 2D structures in Ref.~\cite{jordaan_measuring_2018}, which does not account for Willis coupling. This method makes use of the incident and scattered pressure fields around the object and fits Bessel and Hankel functions to them as $p_\mathrm{inc}(r,\theta) = \sum_n\beta_nJ_n(kr)e^{in\theta}$ and $p_\mathrm{scat}(r,\theta)=\sum_n\gamma_nH^{(1)}_n(kr)e^{in\theta}$. Since only monopole and dipole components are of importance, the problem can be reduced to considering $n=-1, 0,1$ terms. Following this, the incident pressure at the meta-atom center is $\breve{p}^{\mathrm{inc}} = \beta_0$ and the particle velocity $\breve{v}^{\mathrm{inc}}_{x,y} = \frac{\beta_{1} \mp \beta_{-1}}{2 c \rho_0}$. The monopole moment $M = \frac{-4 \gamma_0}{ik^2c^2}$ and the dipole moments $D_{x, y} = \frac{-4 (\gamma_1 \mp \gamma_{-1})}{ik^3c^2}$ can be retrieved from the scattered field.

The expansion coefficients $\beta_n$ and $\gamma_n$ can be obtained from measured or numerically determined pressure on circles with radii $R^{\mathrm{inc}}$ and $R^{\mathrm{scat}}$ respectively. From the orthogonality of exponential functions the coefficients can be found as
\begin{equation} \label{eq:beta_n}
\beta_n = \frac{1}{2 \pi J_n(k R^{\mathrm{inc}})}\int_{0}^{2\pi} p^{\mathrm{inc}}(R^{\mathrm{inc}}, \theta) e^{-in\theta} \mathrm{d}\theta
\end{equation}
and
\begin{equation} \label{eq:gamma_n}
\gamma_n = \frac{1}{2 \pi H_n^{(1)}(k R^{\mathrm{scat}})}\int_{0}^{2\pi} p^{\mathrm{scat}}(R^{\mathrm{scat}}, \theta) e^{-in\theta} \mathrm{d}\theta.
\end{equation}
To avoid the singularity of Eq.~\eqref{eq:beta_n} due to zeros of the Bessel function \cite{jordaan_measuring_2018}, we ensure $R^{\mathrm{inc}}<\frac{2.4}{k}$ . However, $R^{\textrm{scat}}$ should be significantly larger than the meta-atom outer radius to reduce near field contributions. These conflicting requirements mean that $R^{\mathrm{inc}}$ and $R^{\mathrm{scat}}$ are different in general.\\
\\
To fully determine the polarizability in 2D, $\breve{p}^{\mathrm{inc}}$, $\breve{\mathbf{v}}^{\mathrm{inc}}$, $M$ and $\bm{D}$ must be determined for at least 3 incident angles.
The incident field quantities for all available angles $\theta_{1..m}$ are arranged in a  matrix $\bm{\Upsilon}$ as
\begin{equation}
\bm{\Upsilon} =
\begin{bmatrix}
\breve{p}^{\mathrm{inc}}\left(\theta_1\right)	& \breve{v}_x^{\mathrm{inc}}\left(\theta_1\right)	& \breve{v}_y^{\mathrm{inc}}\left(\theta_1\right) \\
\vdots	& \vdots	& \vdots \\
\breve{p}^{\mathrm{inc}}\left(\theta_m\right)	& \breve{v}_x^{\mathrm{inc}}\left(\theta_m\right)	& \breve{v}_y^{\mathrm{inc}}\left(\theta_m\right) \\
\end{bmatrix}
\end{equation}
Knowing $M$ and $\bm{D}$ for each $\theta_{1..m}$, allows the polarizability tensor $\bm{\alpha}$ to be determined by inversion of $\mathbf{\Upsilon}$. For increased robustness, we take additional angles. The polarizability tensor is then determined via least squares as
\begin{equation} \label{eq:alpha_num}
\bm{\alpha}
= \left(\bm{\Upsilon}^T\bm{\Upsilon}\right)^{-1}\bm{\Upsilon}^T
\begin{bmatrix}
M\left(\theta_1\right) & D_x\left(\theta_1\right) & D_y\left(\theta_1\right) \\
\vdots & \vdots & \vdots \\
M\left(\theta_m\right) & D_x\left(\theta_m\right) & D_y\left(\theta_m\right)
\end{bmatrix}.
\end{equation}

\subsection{Numerical model}

To obtain a numerical solution for the polarizability a custom 2D boundary element method code is used, as described in Ref.~\cite{jordaan_measuring_2018}.
It uses continuous elements with a quadratic interpolation functions \cite{marburg_computational_2008} and discretization by collocation method \cite{wu_boundary_2000} with an adaptive integration scheme \cite{marburg_boundary_2018}.
Initially the solids are treated as acoustic hard boundaries, except for the results shown in Fig.~\ref{fig:impedance}, where surface impedance boundary conditions are used. Thermo-viscous losses in air are not included in this formulation. The incident field is a unit intensity plane wave, therefore $\breve{p}^{\mathrm{inc}}$ and $\breve{\mathbf{v}}^{\mathrm{inc}}$ are known explicitly.

To calculate the influence of thermo-viscous losses in air on the polarizability of the meta-atom (Fig.~\ref{fig:lossy}) the commercial \gls{fem} code COMSOL Multiphysics was utilized.

\subsection{Waveguide Scattering Experiment}

To experimentally determine the incident and scattered pressure fields we use a 2D anechoic waveguide chamber presented in \cite{jordaan_measuring_2018}. The propagation medium is air with an acoustic velocity of $\unit[343]{m/s}$. The height of the chamber (\unit[66]{mm}) supports only a single propagation mode at frequencies up to \unit[2600]{Hz}, making it a 2D wave propagation system. The excitation source is a speaker, excited by a continuous wave with frequency varying between \unit[1000]{Hz} and \unit[2000]{Hz} in \unit[40]{Hz} steps. The response is measured by a microphone, which is moved in two axes by belts driven by stepper motors.

The sample is a single aperture meta-atom shown in Fig.~\ref{fig:shape}(a).
This meta-atom was manufactured on a RepRap X400 3D printer using PLA filament with \unit[100]{\%} filling. Additionally, rubber seals (black) are glued on top and bottom to prevent air leakage from the resonator cavity and to achieve more homogeneous clamping.
The incident field is measured at a radius $R^{\textrm{inc}} = \unit[40]{mm}$.
The scattering of the sample is measured at 6 different incident angles ($0^{\circ}$, $90^{\circ}$, $180^{\circ}$, $270^{\circ}$, $60^{\circ}$, $120^{\circ}$) at a radius of $R^{\textrm{scat}} = \unit[200]{mm}$.

\end{document}